\documentclass{elsart}
\usepackage{natbib}
\usepackage{psfig}
\begin{document}
\runauthor{Holt, Tadhunter and Morganti}
\begin{frontmatter}
\title{Emission line studies of young radio sources}
\author[Sheffield,Kapteyn]{Joanna Holt\thanksref{Holt}}
\author[Sheffield]{Clive Tadhunter}
\author[Astron]{Raffaella Morganti}

\address[Sheffield]{Department of Physics \& Astronomy, University of
Sheffield, Sheffield, S3 7RH, UK.}
\address[Kapteyn]{Kapteyn Astronomical Institute, Postbus 800,
NL-9700 AV Groningen, The Netherlands.}
\address[Astron]{Netherlands Foundation for Research in Astronomy,
Postbus 2, NL-7990 AA Dwingeloo, The Netherlands.}
\thanks[Holt]{Email: j.holt@sheffield.ac.uk}
\begin{abstract}
We demonstrate the efficiency of high quality optical spectroscopic
observations of two compact radio sources, PKS 1549-79 and PKS 1345+12,
as a probe of the kinematics and
physical conditions in the circumnuclear gas in the early stages of
radio source evolution. We outline a schematic model for PKS 1345+12
based on the model for PKS 1549-79 proposed by Tadhunter et al. (2001)
in which the young radio source is expanding through a dense and dusty
enshrouding cocoon, sweeping material out of the circumnuclear regions. 
\end{abstract}
\end{frontmatter}

\section{Introduction}
\typeout{SET RUN AUTHOR to \@runauthor}
Gigahertz-Peaked Spectrum Radio Sources (GPS: D $<$ 1 kpc) and the
larger Compact Steep Spectrum Radio Sources (CSS: D $<$ 15 kpc)
account for a 
significant fraction of the radio source population ($\sim$
40\%) though their nature is not fully understood (see O'Dea 1998 
and references therein). Currently, we believe they are {\em young} radio 
sources (Fanti et al. 1995) supported by
estimates of dynamical ages: t$_{dyn}$
$\sim$ 10$^{2}$ - 10$^{3}$ years (Owsianik et al. 1998); and radio spectral
ages: t$_{sp}$ $<$ 10$^{4}$ years (Murgia et al. 1999). This is in
preference to the {\em frustration scenario} where the ISM is so dense,
the radio jets cannot escape and the radio source remains confined and
frustrated for its entire lifetime (van Breugel 1984). 

If compact radio sources are young radio sources, we will
observe them relatively recently after the event(s) which triggered
the activity (e.g. a merger; Heckman et al. 1986).
Indeed, many compact radio sources
exhibit features attributed to mergers such as double nuclei, tidal
tails, arcs of emission and distorted isophotes
(e.g. Stanghellini et al. 1993). Hence, the host galaxy will still retain the
dense ISM injected by the merger. The young radio jets will
also be relatively small and so will readily interact with the dense ISM.

In these proceedings we demonstrate the efficiency of 
high quality optical spectroscopic observations 
as a probe of the kinematics and physical
conditions of the ISM in the host galaxies of radio sources,
particularly in the early stages of radio source evolution. We discuss 
observations of
two compact radio sources, the flat spectrum radio
source PKS 1549-79 and the
GPS source PKS 1345+12, and outline a schematic model for PKS 1345+12.
For detailed discussions see Tadhunter et al. (2001) and Holt et
al. (2002, 2003). 

\section{PKS 1549-79}
\begin{figure*}
\centerline{\psfig{file=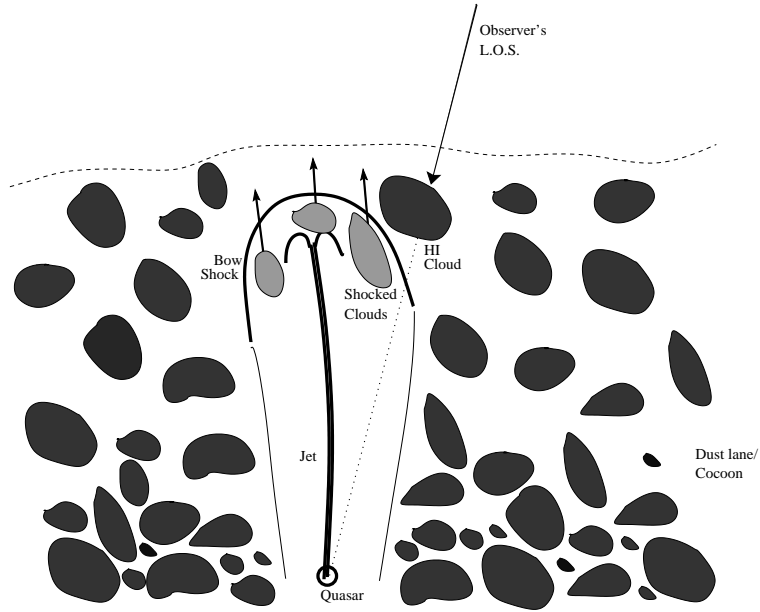,width=10cm,angle=-90.}}
\caption{Schematic model for PKS 1549-79 proposed by Tadhunter et al. (2001).}
\label{Figure 1}
\end{figure*}

From low resolution spectra of the 
compact flat-spectrum radio source PKS 1549-79, 
Tadhunter et al. (2001) report:
\begin{itemize}
\item highly broadened forbidden emission lines (FWHM $\sim$ 1350 km
s$^{-1}$).
\item 2 distinct redshift systems: the high ionisation forbidden lines
(e.g. {[O III]}) are blueshifted by $\sim$ 600 km  s$^{-1}$ with respect
to the low ionisation lines (e.g. {[O II]}) and HI 21 cm
absorption. 
\end{itemize}
Tadhunter et al. (2001) interpret the blueshifted emission lines as
material in outflow.
Due to the small scale radio source ($\sim$ 300 pc), the radio
morphology (core-jet) and the flat radio spectrum, 
they proposed a schematic model in which the young radio source is
enshrouded by a cocoon of dense gas and dust completely obscuring the
central QSO nucleus. The young radio jets are
expanding and sweeping material out of the nuclear regions with the
direction of propagation close to the observer's line-of-sight (see
Fig 1).

\section{PKS 1345+12}
\begin{figure*}
\centerline{\psfig{file=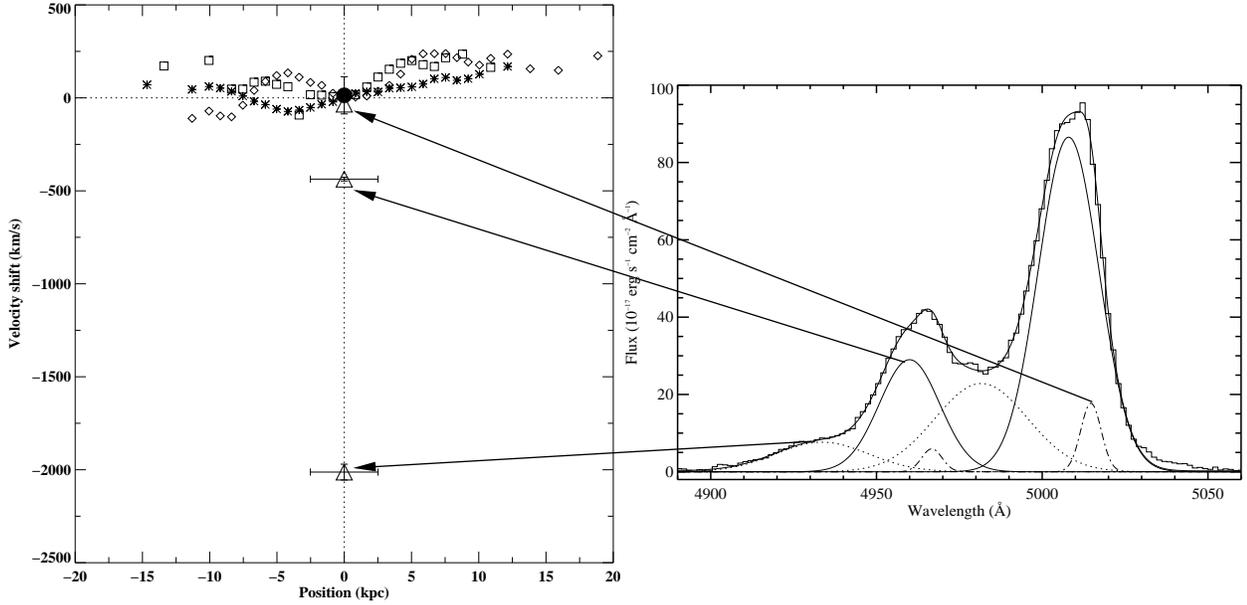,width=17cm}}
\caption{{\em Left}: Radial velocity profile of the extended gaseous
halo. Key: $\ast$, $\Diamond$ and $\Box$: 
highly extended {[O II]} emission (narrowest component); 
$\bullet$: HI 21 cm absorption (Mirabel
1989); $\triangle$: 3 components of {[O III]} in the nuclues, from 
{\em right}).}
\label{Figure 2}
\end{figure*}

We present here a summary of results from intermediate resolution
($\sim$ 4\AA) spectra over a wide spectral range ($\sim$ 4500\AA) from 
ISIS on the 4.2m WHT on La Palma. For a detailed discussion see Holt
et al. (2003).

From modelling of the highly extended {[O
II]}$\lambda\lambda$3726,3729 emission line ($\sim$ 20 kpc) using
between 1 and 3 Gaussian components, we believe that the narrowest of
these represents the galaxy rest frame:
\begin{itemize}
\item It is the only component observed across the entire spatial range;
\item It is consistent with the deep HI absorption
(Mirabel 1989; Morganti et al. 2002) (see Fig 2)
and the stellar absorption lines (Grandi 1977);
\item It is the least kinematically disturbed component (FWHM $\sim$ 340 km
s$^{-1}$);
\item It has a small velocity amplitude ($\Delta_{vel}$ $<$ 250 km
s$^{-1}$; see Fig 2)
consistent with gravitational motions (Tadhunter et al. 1989);
\item It is the only component consistently observed in all emission
lines, irrespective of the model required to reproduce the emission
line profile.
\end{itemize}

In the nuclear aperture, the emission lines 
are broad with strong blue
asymmetries. It is essential to model all emission lines with 
3 Gaussian components
(see Fig 2, also Holt et al. 2003). The
narrowest of these is consistent with the narrow component of {[O II]}
and the rest frame of the galaxy. The intermediate and
broad components are blueshifted by up to $\sim$ 2000 km s$^{-1}$
with respect to this narrow component. Due to large
reddening in the nucleus (see below), we interpret this as
material in outflow. However, one model does not reproduce all
emission lines -- different velocity widths and
shifts for the broad and intermediate components are required for {[O
I]}$\lambda\lambda$6300,6363 and {[S II]}$\lambda\lambda$6716,6731 
(see Holt et al. 2003).

We estimated the density in the nucleus using the density diagnostic,
the {[S II]}$\lambda\lambda$6716,6731 doublet. The intermediate and
broad components have densities consistent with the high density
limit. By varying the fit parameters, we obtained lower limits on on
the density of n$_{e}$ $>$ 5300 cm$^{-3}$ and n$_{e}$ $>$ 4200
cm$^{-3}$ for the intermediate and broad components respectively. The
narrow component is consistent with the low density limit (n$_{e}$ $<$
150 cm$^{-3}$). 

We have investigated reddening in the nucleus of PKS 1345+12 using
three independent techniques, each assuming a simple foreground screen
for interstellar extinction and the Seaton (1979) extinction law. 
\begin{itemize}
\item {\em Balmer line ratios}.
The H$\alpha$/{[N II]} blend is highly complex. 
We measure H$\alpha$/H$\beta$ ratios
of 3.32 $\pm$ 0.33, 5.25 $\pm$ 0.28 and 18.81 $\pm$ 4.74 corresponding
to E(B-V) values of 0.06 $\pm$ 0.05, 0.42 $\pm$ 0.10 and 1.44 $\pm$
0.50 for the narrow, intermediate and broad components
respectively. By varying the model fitting parameters, 
we estimate a lower limit on
the reddening in the broadest component of E(B-V) $>$ 0.92.
\item {\em Comparison with the infra-red}.
Veilleux et al. (1997) present measurements of Pa$\alpha$. The
broadest component is consistent in velocity width and shift with the
broadest component in {[O III]}. By varying the quoted flux of
Pa$\alpha$ by a factor of 2, to account for possible slit loss
differences between the observations, Pa$\alpha$/H$\beta$ gives a
range of E(B-V) values:
1.60 $<$ E(B-V) $<$ 2.00. This is consistent, within the errors,
with the measured value for H$\alpha$/H$\beta$.
\item {\em The nebular continuum}.
Before modelling the faint lines in the nucleus, we subtracted both
the nebular (see Dickson et al. 1995) and stellar
continuum. If zero reddening is assumed for all components, a
large discontinuity, corresponding to the Balmer edge at
$\sim$ 3645 \AA, remains when subtracting the nebular continuum 
from the spectrum (see Fig 7 in Holt et
al. 2003). However, if the measured E(B-V) values are assumed, the
discontinuity disappears. The strength of this discontinuity is
particularly sensitive to the amount of flux in the intermediate
component and we estimate a lower limit of E(B-V) $>$ 0.3 for the
intermediate component.  
\end{itemize}
The nucleus of PKS 1345+12 is highly reddened, with reddening
increasing with broadness of component.

We estimate an upper limit on the mass of line emitting gas
of $<$ 10$^{6}$ M$_{\odot}$ (from the density and reddening-corrected
H$\beta$ luminosity) which is consistent with the radio source 
being young.

\section{The model}
\begin{figure*}
\centerline{\psfig{file=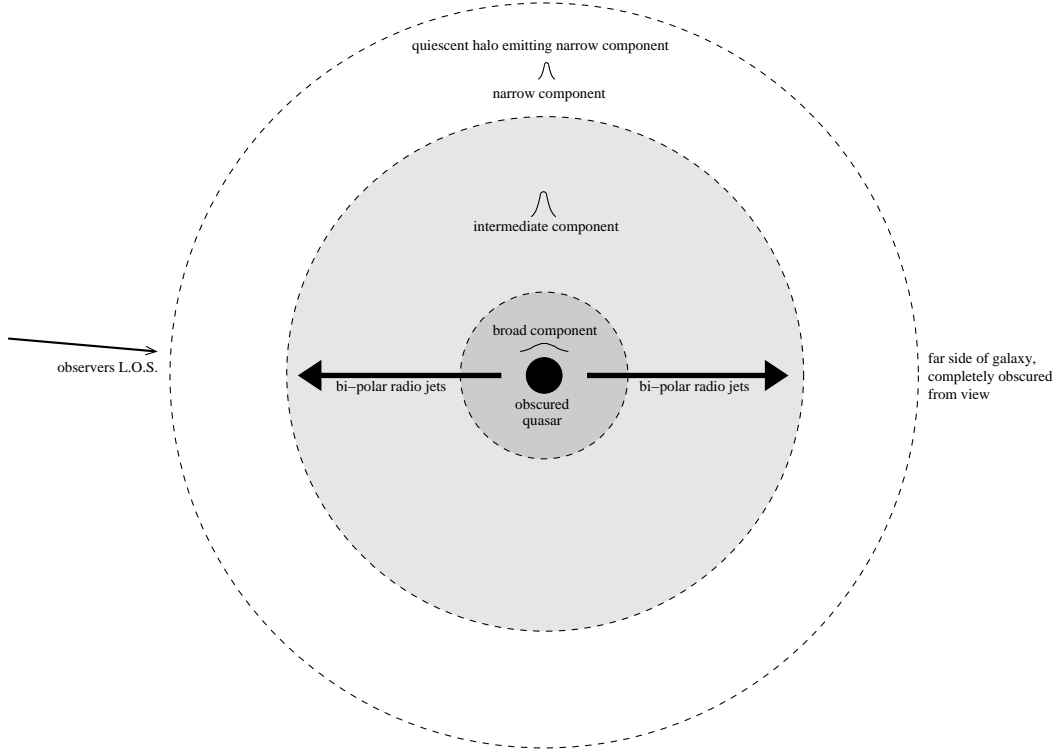,width=14cm}}
\caption{Schematic model for PKS 1345+12.}
\label{Figure 3}
\end{figure*}

Combining these results, we have developed the model from Tadhunter et 
al. (2001). Fig 3 shows the schematic model which retains the
small scale radio jets expanding out through an enshrouding cocoon
giving rise to emission line outflows.
New features include stratification of the ISM -
all emission lines in the nucleus require 3 Gaussian components though 
the velocity width and shift of the intermediate and broad components 
varies between emission lines. However, gradients (e.g. density, velocity,
ionisation potential) must exist across the regions emitting the
intermediate and broad components.
The narrow component remains consistent in
all emission lines and represents the relatively quiescent
halo. `Strata' positioning is determined by the reddening - the
reddening increases with broadness of component and so the broadest
component must originate  from the region closest to the nucleus.

\section{Acknowledgements}
JH acknowledges a PPARC PhD studentship and a Marie Curie Fellowship. 
The William Herschel Telescope is operated on the
island of La Palma by the Isaac Newton Group in the Spanish
Observatorio del Roque de los Muchachos of the Instituto de
Astrofisica de Canarias.

\end{document}